\documentclass[aps,prb,twocolumn,floatfix,citeautoscript,nofootinbib,superscriptaddress]{revtex4-2}
\usepackage{amsbsy}
\usepackage{latexsym,epsfig,graphicx}
\usepackage{dcolumn}
\usepackage{graphicx}
\usepackage{subfigure}
\usepackage{comment}
\usepackage{color}
\usepackage{xcolor}
\usepackage{soul}
\usepackage{bm}
\usepackage{mathrsfs}
\usepackage{amsfonts}
\usepackage{amsmath}
\usepackage{amssymb}
\usepackage{xspace}
\usepackage{epstopdf}
\usepackage{tabularx}
\usepackage{longtable}
\usepackage[colorlinks=true, letterpaper=true, pdfstartview=FitV, linkcolor=blue, citecolor=blue, urlcolor=blue]{hyperref}
\usepackage[normalem]{ulem}

\makeatletter
\newsavebox{\@brx}
\newcommand{\llangle}[1][]{\savebox{\@brx}{\(\m@th{#1\langle}\)}%
  \mathopen{\copy\@brx\kern-0.5\wd\@brx\usebox{\@brx}}}
\newcommand{\rrangle}[1][]{\savebox{\@brx}{\(\m@th{#1\rangle}\)}%
  \mathclose{\copy\@brx\kern-0.5\wd\@brx\usebox{\@brx}}}
\makeatother

\begin{document}

\title{Quantized Quadrupole Superconductors}
\author{Yun-Mei Li}
\email[contact author:]{yunmeili@zju.edu.cn}
\affiliation{Center for Quantum Matter, School of Physics, Zhejiang University, Hangzhou 310027, China}
\author{Yongwei Huang}
\affiliation{School of Physics, Ningxia University, Yinchuan 750021, China}
\author{Kai Chang}
\email[contact author:]{kchang@zju.edu.cn}
\affiliation{Center for Quantum Matter, School of Physics, Zhejiang University, Hangzhou 310027, China}

\begin{abstract}
  We introduce a class of superconductors termed ``quantized quadrupole superconductors" that support Majorana corner modes according to the bulk-corner correspondence,
  distinct from previous works on the second-order topological superconductors. An intrinsic physical quantity for superconductors, i.e., the quadrupole moment serves
  as the topological invariant, which is always half-quantized due to the particle-hole symmetry.
  As examples, two types of mixed pairings, $d_{x^{2}-y^{2}}\pm i d_{xy}$ and $d_{x^{2}-y^{2}}\pm i s$,
  induced in the bilayer two-dimensional electron gases with Rashba spin-orbit coupling give the quadrupole phase.
  Extended discussions indicate that the nontrivial phase is robust against relative phase fluctuations in the mixed pairings and the disorders.
  Our schemes provide realistic platforms to implement Majorana zero modes, paving the way for studying the Majorana physics.
\end{abstract}

\maketitle

\section{Introduction}
Topological superconductors (TSCs) support Majorana excitations obeying non-Abelian exchange statistics,
which are the key ingredient in topological quantum computations~\cite{AKitaev,MZHasan,XLQi,SRElliott,AKitaev2,CNayak}.
One of the key features in TSCs is the bulk-boundary correspondence~\cite{MZHasan,XLQi}.
Usually, a nontrivial topological invariant derived from the bulk states under the assumptions
of periodic boundary conditions (PBC) guarantees the emergence of localized Majorana modes
at the boundaries (ends, edges or surfaces) when cutting the periodic systems along one certain direction~\cite{AKitaev,XLQi,JAlicea,TDStanescu,CBeenakker,CZhang,JAlicea,TDStanescu,CBeenakker,YOreg,RMLutchyn,STewari,SNakosai,FZhang}.
The recognition of higher-order topology is expected to extend or enrich this correspondence~\cite{WABenalcazar1,WABenalcazar2,JLangbehn,ZSong,MGeier,YMLi,EKhalaf}.
The Majorana excitations appear at the intersections (corners or hinges) of adjacent boundaries
when cutting the periodic system along two or three different directions in higher-order TSCs~\cite{EKhalaf,MGeier,ZYan,QWang,XZhu1,YVolpez,XZhu2,RXZhang,YWu,AChew,ZYan2,YXLi,YTHsu,AKGhosh}.
Part of the works utilize the crystalline symmetry analysis for the topological characterizations~\cite{JLangbehn,ZSong,MGeier,EKhalaf,ZYan2}.
However, this approach can not explain the persistence of the higher-order states when the related crystalline symmetries are broken.
The others aim at gapping the first-order edge or surface states by in-plane magnetic fields~\cite{XZhu1,YVolpez,YWu,YVolpez,YXLi,AKGhosh}
or unconventional superconductors~\cite{ZYan,QWang,XZhu2,RXZhang,YTHsu,ZYan2}.
The recent works~\cite{HWang,XZhu3} attempt to connect the appearance of the Majorana corner modes to the bulk spectrum. However,
the higher-order TSCs with bulk-boundary correspondence still remain unexplored.

In this Letter, we propose a class of superconductors dubbed as ``quantized quadrupole superconductors" that support Majorana corner modes based on
bulk-corner correspondence. By generalizing the quadrupole moment to the superconductors, we find that it is always half-quantized guaranteed only by
the particle-hole symmetry in superconductors and thus behaves as an intrinsic physical quantity,
different from that in electronic systems relying on the crystalline symmetries.
A one-half quadrupole moment implies the emergence of zero-energy corner modes in a square or rectangle sample under the open boundary condition (OBC), demonstrated
to be Majorana corner modes (MCMs), also guaranteed by the particle-hole symmetry.
Distinct from the previous works, our approach to the second-order TSCs only utilizes the particle-hole symmetry in superconductors, unique
with requiring the minimal assumptions or conditions and revealing a new class of superconductors.

As concrete examples, we show that the bilayer spin-orbit coupled two-dimensional electron gases with at least two types of proximity pairings can host nontrivial quadrupole phase. The first pairing is the $d_{x^{2}-y^{2}}\pm id_{xy}$ ($d\pm id^{\prime}$ for short) while the second is $d_{x^{2}-y^{2}}\pm is$ ($d\pm is$ for short).
We take the twisted bilayer copper oxides to induce the $d\pm id^{\prime}$ for calculations~\cite{YYu,OCan}.
The $d\pm is$ pairing can be realized by the Josephson junctions~\cite{RKleiner,PVKomissinski,PKomissinskiy,SCharpentier}.
In the region where quadrupole moment does not vanish, there emerges
single Majorana corner mode (MCM) at each corner under OBC.
In the trivial region, the sample supports quasiparticle corner modes or no corner modes.
The bulk gap closing and reopening give the phase transitions between the nontrivial and trivial region.
The MCMs and bulk-corner correspondence in quadrupole superconductors are robust against the disorders.
The quadrupole phase also survives from a large phase fluctuations in the mixed pairing.
Besides, quadrupole superconductors are intrinsic higher-order TSCs, providing
feasible schemes for experimentalists to achieve MCMs.

\section{The bulk quadrupole moment in superconductors}
We establish the general theory of quantized quadrupole superconductors firstly.
By generalizing Resta formula for electric polarization~\cite{RResta} and seminal works on the electric multipole insulators~\cite{WAWheeler,BKang},
the quadrupole moment for superconductors is defined in the real-space as
\begin{equation}\label{eq1}
  q_{xy}=\frac{1}{2\pi}\mathrm{Im}\log[\det(\langle U|\hat{Q}|U\rangle)].
\end{equation}
$\hat{Q}=e^{2\pi i\hat{x}\hat{y}/(L_{x}L_{y})}$. $\hat{x}$ ($\hat{y}$) is the position operator along the $x$ ($y$) dimension and $L_{x,y}$ the corresponding size.
The matrix $U$ is constructed by the quasiparticle states of negative energy branches obtained by diagonalizing
the Bogoliubov-de Gennes (BdG) Hamiltonian. The quasiparticle eigenstates are defined on the torus with PBC
such that only the bulk states determine the properties of $q_{xy}$.
There is a reasonable rule on the operator $\hat{Q}$.
For any spin, orbital and particle-hole degree of freedom on each site or coordinate, the position operator is adopted to be the same.
When considering the spin and particle-hole degree of freedom, $\det(\hat{Q})=1$ in usual cases~\cite{SM}.

The particle-hole symmetry in superconductors guarantees that
$q_{xy}$ is always quantized to $0$ or $\frac{1}{2}$ because $\det(U^{\dagger}\hat{Q}U)$ is real.
The details are presented in the Supplementary Materials (SMs)~\cite{SM}.
We use  Sylvester's determinant identity $\det(\mathbf{1}+AB)=\det(\mathbf{1}+BA)$ for the derivations.
$\det(U^{\dagger}\hat{Q}U)=\det[U^{\dagger}(\hat{Q}-\mathbf{1}+\mathbf{1})U]=\det[\mathbf{1}+(\hat{Q}-\mathbf{1})UU^{\dagger}]$.
Let $V$ be the quasiparticle states of the positive energy branch.
From the relation $UU^{\dagger}+VV^{\dagger}=\mathbf{1}$, we have
\begin{equation}
   \det[\mathbf{1}+(\hat{Q}-\mathbf{1})(\mathbf{1}-VV^{\dagger})] = \det(V^{\dagger}\hat{Q}^{\dagger}V).  \label{eq2}
\end{equation}
Eq.~\eqref{eq2} implies that $\det(U^{\dagger}\hat{Q}U)=\det(V^{\dagger}\hat{Q}^{\dagger}V)$.
The particle-hole symmetry is usually of the form $\mathcal{P}=\tau_{x}\mathcal{K}$, relating $U$ and $V$ as $V=\mathcal{P}U=\tau_{x}U^{*}$,
where $\tau_{x}$ acts on the particle-hole degree of freedom and $\mathcal{K}$ denote the complex conjugation.
Then $\det(V^{\dagger}\hat{Q}^{\dagger}V)=\det(U^{\textrm{T}}\tau_{x}\hat{Q}^{\dagger}\tau_{x}U^{*})
  =[\det(U^{\dagger}\hat{Q}U)]^{*}$, indicates the half-quantization of  $q_{xy}$.
We here restrict $2q_{xy}=1$ mod $2$ to make $q_{xy}=0$ or $\frac{1}{2}$.

The properties of the quadrupole moment for the two-dimensional (2D) superconductors are quite different from that for electronic systems.
The half-quantization of $q_{xy}$ in electronic systems~\cite{WAWheeler,BKang,SOno,YTada}
relies on the mirror, four-fold rotational or chiral symmetries, which can be broken by the perturbations, such as the disorders, strains and other symmetry breaking terms.
By contrast, the quadrupole moment for 2D superconductors behaves as an intrinsic physical quantity as its half-quantization is protected by the always present particle-hole symmetry in superconductors
and is independent on the crystalline symmetries including the inversion symmetry~\cite{EKhalaf}.
The superconductors with $q_{xy}=\frac{1}{2}$ are expected to be robust against the above perturbations.
More importantly, when $q_{xy}=\frac{1}{2}$,
$\partial_{x}\partial_{y}q_{xy}$ is not zero only at the corners of a square or rectangle shaped sample under OBC, indicating
the emergence of single zero-energy mode at each corner. These zero-energy states can be expressed as $\psi=\sum_{i}(u_{i}c_{i}+v_{i}c_{i}^{\dagger})$, where $i$ denote the combined orbital, spin and lattice index.
Due to the particle-hole symmetry, we have $\psi=\mathcal{P}\psi=\sum_{i}(v_{i}^{*}c_{i}+u_{i}^{*}c_{i}^{\dagger})$, making the relation $v_{i}=u_{i}^{*}$
and $\psi^{\dagger}=\psi$. The spatially separated zero-energy corner modes are thus Majorana corner modes.
The quantized quadrupole superconductors then exhibit a bulk-corner correspondence.

\section{Realistic examples for quantized quadrupole superconductors}
\subsection{The electronic Hamiltonian} 
\begin{figure}[t]
  \centering
  \includegraphics[width=0.5\textwidth]{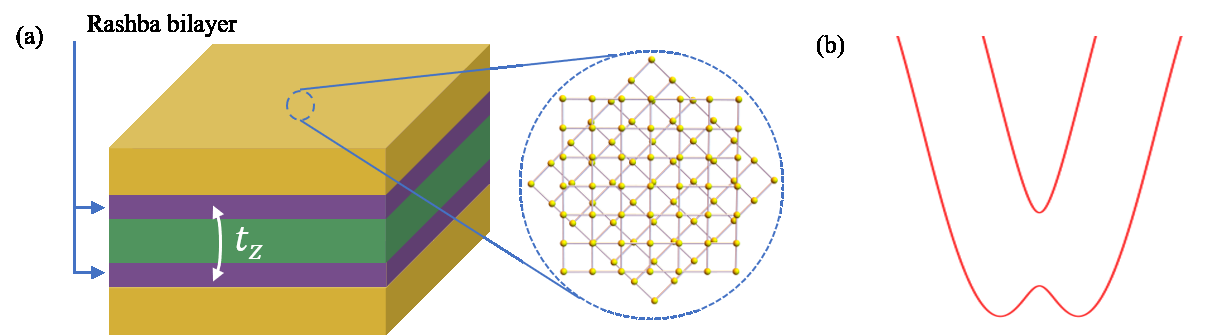}\\
  \caption{(a) The illustration for the setup to achieve the first model for quadrupole superconductor that supports Majorana corner modes.
  The bilayer two-dimensional electron gases with Rashba spin-orbit coupling are in proximity to twisted bilayer copper oxides.
  The second model is by replacing the twisted bilayer copper oxides with Josephson junctions displaying $d\pm is$ pairings.
  (b) The bands for the bilayer 2DEGs near $\Gamma$ point. The interlayer coupling $t_{z}$ splits the two Rashba bands into bonding (lower)
   and antibonding states with a hybridization gap.}\label{fig1}
\end{figure}

To obtain nontrivial quadrupole superconductors, the single-particle system is the
coupled bilayer two-dimensional electron gases (2DEGs) with Rashba spin-orbit coupling (SOC), as illustrated
in Fig.~\ref{fig1} (a). We note that the SOC in the bilayer 2DEGs are naturally opposite to each other
due to the opposite structural asymmetry between the bottom and top layers. We employ a tight-binding Hamiltonian on a square lattice
to describe the bilayer 2DEGs,
\begin{equation}\label{eq3}
  H_{0}=\sum_{\langle ij\rangle,ss^{\prime}\sigma}c_{i\sigma s}^{\dagger}[-t_{0}\delta_{ss^{\prime}}
  +i\alpha\sigma_{z}(\mathbf{s}\times\hat{\mathbf{d}}_{ij})_{z}]c_{j\sigma s^{\prime}}+H_{z},
\end{equation}
where $H_{z}=t_{z}\sum_{is,\sigma\neq\sigma^{\prime}}c_{i\sigma s}^{\dagger}c_{i\sigma^{\prime} s}$
is the tunneling term between the two layers.
$\hat{\mathbf{d}}_{ij}$ is the unit vector pointing from site $i$ to site $j$.
$\mathbf{s}$ and $\bm\sigma$ denote the spin and layer indices, respectively.
$\alpha$ characterizes the strength of the SOC. With a Fourier transformation to the momentum space,
$H_{0}=\sum_{\mathbf{k}}\psi_{\mathbf{k}}^{\dagger}h_{\mathbf{k}}\psi_{\mathbf{k}}$
in the basis of $\psi_{\mathbf{k}}=(c_{\mathbf{k},1\uparrow},c_{\mathbf{k},1\downarrow},c_{\mathbf{k},2\uparrow},c_{\mathbf{k},2\downarrow})^{\textrm{T}}$,
where
\begin{equation}\label{eq4}
  h_{\mathbf{k}}=\xi_{\mathbf{k}}+2\alpha(s_{x}\sin k_{y}-s_{y}\sin k_{x})\sigma_{z}+t_{z}\sigma_{x},
\end{equation}
and $\xi_{\mathbf{k}}=-2t_{0}(\cos k_{x}+\cos k_{y}-2)$. The $2$ in the bracket of $\xi_{\mathbf{k}}$ is for resetting the zero energy position.
The interlayer coupling hybridize the two Rashba bands at the top and bottom layers with a gap
$2t_{z}$ at $\Gamma$ [$\mathbf{k}=(0,0)$] and $\mathrm{M}$ [$\mathbf{k}=(\pi,\pi)$] points, as illustrated in Fig.~\ref{fig1} (b).

\subsection{$d\pm id^{\prime}$ proximity pairing}
There are many ways to induce the $d\pm id^{\prime}$ proximity paring in the 2DEGs~\cite{OCan,RBLaughlin, ZYang}.
We choose the twisted bilayer copper oxides proposed in the recent work~\cite{OCan} to proximitize the 2DEGs inducing our desired mixed pairing, as illustrated in Fig.~\ref{fig1} (a).
One of the candidate materials is the van der Waals-bonded high-critical-temperature copper oxide materials,
such as Bi$_{2}$Sr$_{2}$CaCu$_{2}$O$_{8+\delta}$ (Bi2212), which gives $d_{x^{2}-y^{2}}$ pairing upon a high critical temperature $T_{c}\simeq 90$ K.
When the twist angle $\theta\simeq 45^{\circ}$, the relative phase between two layers is about $\pm\pi/2$ to generate $d\pm id^{\prime}$ pairing.
The relative phase would spontaneously choose one of the two values~\cite{OCan}.
For simplicity, we here restrict the bottom (top) 2DEG layer are in proximity contact to
the twisted bilayer cooper oxides with the relative phase $\pi/2$ ($-\pi/2$) or vice versa,
achievable experimentally by proper choosing the twisted samples or phase locking techniques.
The case for other twist angles and relative phases will be discussed at the end.
We can express the proximity superconducting pairing in the Rashba bilayer as
\begin{eqnarray}
   H_{\Delta}^{a} &=& \sum_{\sigma=1}^{2}[ i\Delta^{\prime}\sigma_{z}^{\sigma\sigma}(\sum_{\llangle ij\rrangle_{\hat{1}}}c_{i\sigma\uparrow}^{\dagger}c_{j\sigma\downarrow}^{\dagger}-
  \sum_{\llangle ij\rrangle_{\hat{2}}}c_{i\sigma\uparrow}^{\dagger}c_{j\sigma\downarrow}^{\dagger}) \nonumber \\
  &&+\Delta(\sum_{\langle ij\rangle_{\hat{x}}}c_{i\sigma\uparrow}^{\dagger}c_{j\sigma\downarrow}^{\dagger}-
  \sum_{\langle ij\rangle_{\hat{y}}}c_{i\sigma\uparrow}^{\dagger}c_{j\sigma\downarrow}^{\dagger})+H.c.] ,  \label{eq5}
\end{eqnarray}
$\langle ij\rangle_{\hat{x}(\hat{y})}$ denotes the nearest neighbor along the $x$ ($y$) direction.
$\llangle ij\rrangle_{\hat{1}}$ ($\llangle ij\rrangle_{\hat{2}}$)
denotes the second nearest neighbor along the $\hat{x}+\hat{y}$ ($\hat{x}-\hat{y}$) direction.

The total Hamiltonian in the BdG formalism under the basis $\Psi_{\mathbf{k}}=(\psi_{\mathbf{k}}, \psi_{-\mathbf{k}}^{\dagger})^{\textrm{T}}$
is given by
\begin{equation}\label{eq6}
  H_{\mathrm{BdG}}^{a}(\mathbf{k})=\left(\begin{array}{cc}
  h_{\mathbf{k}}-\mu  & \Delta_{\mathbf{k}a} \\
  \Delta_{\mathbf{k}a}^{\dagger} & -h_{-\mathbf{k}}^{*}+\mu
  \end{array}\right),
\end{equation}
where $\mu$ is the chemical potential and $\Delta_{\mathbf{k}a}=i\Delta_{\mathbf{k}}^{1}s_{y}\sigma_{0}-\Delta_{\mathbf{k}}^{2}s_{y}\sigma_{z}$
with $\Delta_{\mathbf{k}}^{1}=2\Delta(\cos k_{x}-\cos k_{y})$ and $\Delta_{\mathbf{k}}^{2}=4\Delta^{\prime}\sin k_{x}\sin k_{y}$.
The doubly degenerate quasiparticle spectrum $E=\pm\sqrt{A\pm 2\sqrt{B}}$ with
$A=(\xi_{\mathbf{k}}-\mu)^{2}+4\alpha^{2}(\sin^{2}k_{x}+\sin^{2}k_{y})+t_{z}^{2}+(\Delta_{\mathbf{k}}^{1})^{2}+(\Delta_{\mathbf{k}}^{2})^{2}$
and $B=(\xi_{\mathbf{k}}-\mu)^{2}[t_{z}^{2}+4\alpha^{2}(\sin^{2}k_{x}+\sin^{2}k_{y})]+
t_{z}^{2}(\Delta_{\mathbf{k}}^{2})^{2}$.We can see the bulk spectrum is fully gapped except
when $\mu=\pm t_{z}, 8t_{0}\pm t_{z}$, which will be shown later, are the phase transition points.
The system holds no time-reversal symmetry (TRS)
but preserves a particle-hole symmetry $\mathcal{P}=\tau_{x}\mathcal{K}$:
$\mathcal{P}H_{\mathrm{BdG}}^{a}(\mathbf{k})\mathcal{P}^{-1}=-H_{\mathrm{BdG}}(-\mathbf{k})$.
A nontrivial value of $q_{xy}$ demands gapped bulk and edge spectrum simultaneously.

\begin{figure}[t]
  \centering
  \includegraphics[width=0.48\textwidth]{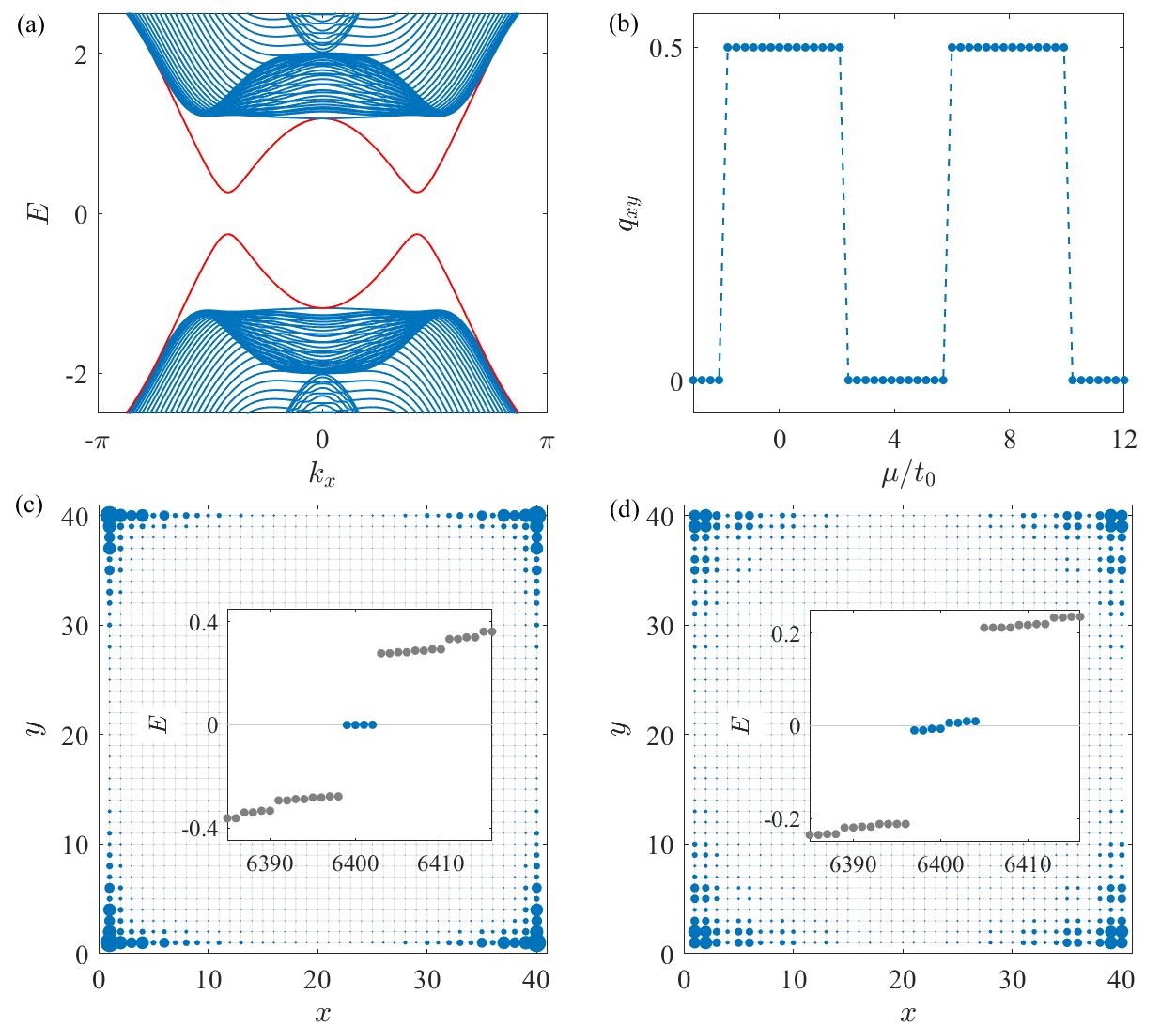}\\
  \caption{(a) The quasiparticle dispersions for a ribbon along the $x$ direction when the Fermi energy $\mu=0$.
  The ribbon width is adopted $N_{y}=40$. (b) The quadrupole moment $q_{xy}$ with respect to $\mu$. The sample size is chosen as $40\times40$ for the calculations.
  (c)-(d) Density distribution for the corner modes with the two chemical potential $\mu=0$ and $\mu=4.0$, respectively.
  The insets show the energy levels around the zero energy. The sample size is $40\times40$.
  In all the panels, $t_{0}=1.0$, $\alpha=1.0$, $t_{z}=2.0$, $\Delta=0.5$, $\Delta^{\prime}=0.4$.
  }\label{fig2}
\end{figure}

We employ the tight-binding Hamiltonian presented in Eq.~\eqref{eq2} and Eq.~\eqref{eq4}
to calculate the quasiparticle energy spectrum for one-dimensional ribbon
to check the existence of edge gap. The quasiparticle bands for a $x$-directional ribbon are
shown in Fig.~\ref{fig2} (a) with chemical potential $\mu=0$. A clear edge gap can be seen.
At other chemical potentials the edge gap always presents even when the
bulk gap closes at $\mu=\pm t_{z}, 8t_{0}\pm t_{z}$~\cite{SM}.
The quadrupole moment $q_{xy}$ with respect to the chemical potential $\mu$ is shown in Fig.~\ref{fig2} (b).
In the ranges $-t_{z}<\mu<t_{z}$ and $8t_{0}-t_{z}<\mu<8t_{0}+t_{z}$,
$q_{xy}$ is $1/2$. In these ranges,
the Fermi level only crosses the bonding or antibonding bands.
The phase transitions are accompanied by the bulk gap closing and reopening.

To demonstrate the bulk-corner correspondence, we calculate quasiparticle energy level
in a square-shaped sample at different $\mu$ under OBC. The results are shown in Figs.~\ref{fig2} (c)
and \ref{fig2} (d). When $\mu$ is in the region giving nontrivial $q_{xy}$,
four MCMs localize at the four corners separately [Fig.~\ref{fig2} (c)].
Due to the high critical temperature superconductors we adopted  ($T_{c}\simeq 90$ K for Bi2212),
these Majorana corner modes are expected to survive up to  a very high temperature.
When $\mu$ locates in the trivial region,
we find eight quasiparticle corner modes with nonzero energy.
There are only four independent physical corner modes while the other four are related by the particle-hole symmetry [Fig.~\ref{fig2} (d)].
The energies of these corner modes do not approach to zero
at sufficiently large sample size~\cite{SM}, demonstrated that they are not MCMs.

\subsection{$d\pm is$ proximity pairing}
\begin{figure}[t]
  \centering
  \includegraphics[width=0.48\textwidth]{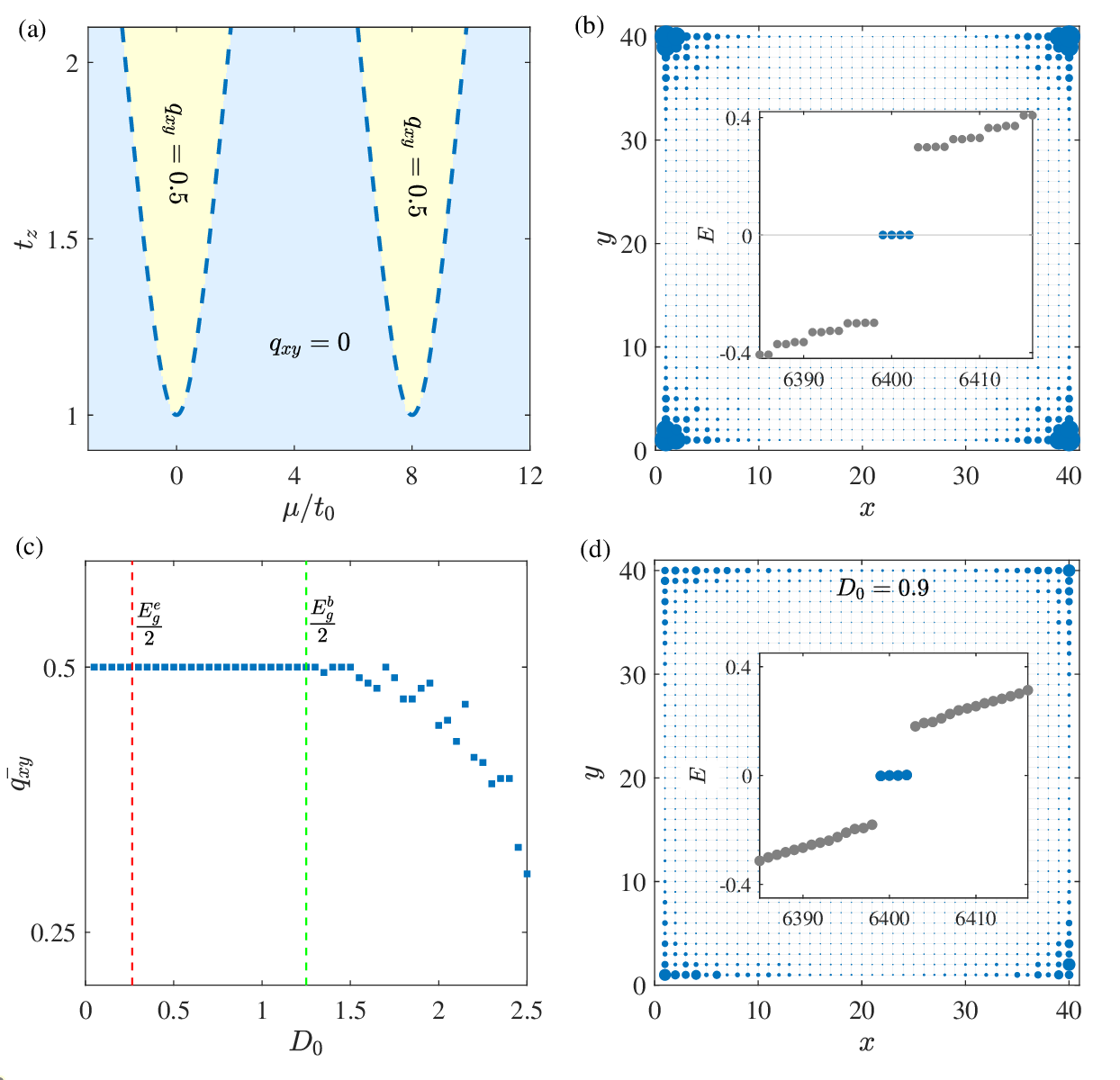}\\
  \caption{(a) The phase diagram in the $t_{z}-\mu$ plane. The sample size is adopted as $20\times20$ for the numerical calculations of $q_{xy}$.
   The dashed lines are the phase boundary obtained analytically from the band gap closing condition. (b) Density distribution of the MCMs with $\mu=0$ and $t_{z}=2.0$.
   The inset shows the energy levels around the zero energy. The sample size is adopted as $40\times40$.
   (c) The averaged $q_{xy}$ with respect to the disorder strength $D_{0}$ for the first model. $E_{g}^{e}$ denotes the edge gap while $E_{g}^{b}$ denotes the bulk gap of quasiparticles.
   (d) The spatial distributions of the MCMs at disorder strength $D_{0}-0.9$ for the first model.}\label{fig3}
\end{figure}

Another type of proximity pairing, i.e., $d\pm is$ can also induce a nontrivial quadrupole phase.
We replace the twisted bilayer copper oxides with a hybrid Josephson junctions consisting of cuprate superconductors and conventional s-wave superconductors.
We here suppose a relative $\pm\frac{\pi}{2}$ phase between the two different pairings.
Similarly, we restrict the bottom (top) 2DEG layer to acquire a $d+is$ ($d-is$) proximity superconducting pairing.
The induced proximity interaction in 2DEGs can be described as
\begin{eqnarray}
  H_{\Delta}^{b} &=& \sum_{\sigma=1}^{2}[\Delta(\sum_{\langle ij\rangle_{\hat{x}}}c_{i\sigma\uparrow}^{\dagger}c_{j\sigma\downarrow}^{\dagger}-
  \sum_{\langle ij\rangle_{\hat{y}}}c_{i\sigma\uparrow}^{\dagger}c_{j\sigma\downarrow}^{\dagger}) \nonumber \\
  && +i\Delta_{0}\sigma_{z}^{\sigma\sigma}\sum_{i}c_{i\sigma\uparrow}^{\dagger}c_{i\sigma\downarrow}^{\dagger}+H.c.].  \label{eq7}
\end{eqnarray}
The BdG formalism of  Hamiltonian can be obtained by replacing $\Delta_{\mathbf{k}a}$ in Eq.~(\ref{eq6}) with
$\Delta_{\mathbf{k}b}=2i\Delta(\cos k_{x}-\cos k_{y})s_{y}\sigma_{0}-\Delta_{0}s_{y}\sigma_{z}$.
The quasiparticle eigenvalues $E=\pm\sqrt{A^{\prime}\pm 2\sqrt{B^{\prime}}}$ with
$A^{\prime}=(\xi_{\mathbf{k}}-\mu)^{2}+4\alpha^{2}(\sin^{2}k_{x}+\sin^{2}k_{y})+t_{z}^{2}+(\Delta_{\mathbf{k}}^{1})^{2}+\Delta_{0}^{2}$
and $B^{\prime}=(\xi_{\mathbf{k}}-\mu)^{2}[t_{z}^{2}+4\alpha^{2}(\sin^{2}k_{x}+\sin^{2}k_{y})]+
t_{z}^{2}\Delta_{0}^{2}$. The quasiparticle gap closes when $t_{z}=\pm\sqrt{\mu^{2}+\Delta_{0}^{2}}$ or $t_{z}=\pm\sqrt{(\mu-8t_{0})^{2}+\Delta_{0}^{2}}$.
The particle-hole symmetry $\mathcal{P}=\tau_{x}\mathcal{K}$ is preserved but the TRS is also broken.
We present the phase diagram, i.e., quadrupole moment $q_{xy}$ with respect to $t_{z}$ and $\mu$ in Fig.~\ref{fig3} (a).
In the regions $|t_{z}|>\sqrt{\mu^{2}+\Delta_{0}^{2}}$
and $|t_{z}|>\sqrt{(\mu-8t_{0})^{2}+\Delta_{0}^{2}}$, $q_{xy}$ is $1/2$, indicating the emergence of the MCMs.
In real systems, we should consider the $\mu$ dependence of proximity pairing~\cite{AMBlackSchaffer}.
Simply replacing $\Delta_{0}$ with $\Delta_{0}(\mu)$ could simulate the realistic situations.
The energy levels of a squared sample and the density distribution of the zero-energy corner states
are shown in Fig.~\ref{fig3} (b).

\section{Release of experimental conditions} 
The recent experiments have successfully fabricated the twisted bilayer cuprates~\cite{YZhu2,SYFrankZhao}
with multiple twist angles. The phenomena related with the TRS breaking are observed.
The hybrid Josephson junctions consisting of cuprate superconductors and conventional $s$-wave superconductors
have also been successfully fabricated experimentally and well studied decade ago~\cite{PVKomissinski,RKleiner,PKomissinskiy}.
The two setups we proposed are experimentally achievable.
We adopt the specific $\pm\frac{\pi}{2}$ relative phase to simplify our calculations and discussions.
We show that the nontrivial quadrupole phase survives to a very large range of the relative phase.

As discussed in Ref.~\cite{OCan}, gapped superconducting phase survives between the critical values of the twist angle of the bilayer cuprate superconductors.
For the first setup, the requirements are that the twist angle is also between the critical values $\theta_{c}^{-}<\theta<\theta_{c}^{+}$
and the sign of the relative phase for the bottom and top bilayer cuprates are opposite.
The twist angles and the pairing amplitudes for the top and bottom bilayer cuprates do not need to be same.
To perform the calculations, we firstly express
the proximity pairing term in a continuous form for any angle,
$\Delta_{\mathbf{k}}^{\sigma}=\Delta(k_{y}^{2}-k_{x}^{2})+\Delta^{\prime}e^{i\varphi (\theta_{\sigma})}[(k_{y}^{\prime})^{2}-(k_{x}^{\prime})^{2}]$,
where $(k_{x}^{\prime},k_{y}^{\prime})^{\mathrm{T}}=R(\theta)(k_{x},k_{y})^{\mathrm{T}}$ and $R(\theta)$ is the 2D representation of SO(2) rotation group.
The phase $\varphi (\theta)$ can be calculated from the twist angle and material parameters~\cite{OCan,SM}.
Then we map $\Delta_{\mathbf{k}}^{\sigma}$ on the square lattice.
The detailed calculations are presented in the SMs~\cite{SM}, presenting same results to Fig.~\ref{fig2} even we adopt different twist angles between top and bottom bilayer.
For the second setup, the nontrivial quadrupole phase survives at a large area of the relative phase between the $d$-wave and $s$-wave pairings.
We present the detailed results in the SMs~\cite{SM}.

Besides, we would like to mention that the bilayer 2DEGs can be replaced by a thin film of 3D strong topological insulators.
The surface states on opposite surface have opposite helicity, perfectly coincides with the Rashba bilayer system.
Our proposal does not require precise square sample, either. See SMs for the MCMs in other geometries~\cite{SM}.

\section{Disorders} 
We turn to the disorder effects. The disorders break all the crystalline symmetries, including the mirror, rotational and inversion symmetries.
 There are at least two sources of disorders, coming from the single-particle and the pairings, respectively.
We take the first model for the discussions. The phase diagram in Fig.~\ref{fig2} (b) depends only on the chemical potential.
The nontrivial phase is robust against on the proximity pairing amplitude fluctuations. Meanwhile,
we do not need to consider the chemical dependence of the proximity pairing amplitude.

For the disorders coming from the electronic states, we consider the random on-site disorders.
The disorder Hamiltonian reads
\begin{equation}\label{eq8}
  H_{dis}=\sum_{i\sigma s}c_{i\sigma s}^{\dagger}V_{i\sigma}c_{i\sigma s}.
\end{equation}
The random potential $V_{i\sigma}$ obeys a Gaussian distribution,
namely $\langle V_{i\sigma}\rangle=0$, $\langle V_{i\sigma}V_{j\sigma^{\prime}}\rangle=D_{0}^{2}\delta_{ij}\delta_{\sigma\sigma^{\prime}}$,
with $D_{0}$ characterizing the strength of the random disorders.
Since the $q_{xy}$ is still half-quantized in the presence of the disorders, we apply 100 different disorder configurations to average the $q_{xy}$ to study the robustness of the MCMs.
The results are presented in Fig.~\ref{fig3} (c). The $q_{xy}$ is robust even when the disorder strength is much larger than the edge gap.
Only when $2D_{0}$ is larger than the bulk gap, the disorders could destroy the bulk-corner correspondence and the MCMs. We present the MCMs in Fig.~\ref{fig3} (d) under a strong disorder configuration.
These behaviors are natural result due to the bulk protection.

\section{Discussions and summary}
We only discuss the quantized quadrupole phase in 2D systems.
Direct extension to the 3D systems could predict the quantized octupole superconductors.
The octupole moment can be defined similar to Eq.~\eqref{eq1}, with $\hat{Q}=e^{2\pi i\hat{x}\hat{y}\hat{z}/(L_{x}L_{y}L_{z})}$.
The realistic systems that exhibit the octupole phase will be discussed in the future.

In summary, we established the higher-order topological superconductors protected by bulk quadrupole moment.
These superconductors are robust against the phase fluctuations of the mixed pairing in realistic systems and also the disorders.
Together with the intrinsic nature, the models we proposed are experimentally feasible by current techniques,
providing alternative platforms for the study of the Majorana excitations and non-Abelian statistics in condensed matter systems.  \\

\section*{Acknowledgements}
Y.-M. Li thanks Prof. Xiancong Lu and Prof. Wei-Nan Lin for their helpful discussions.
Y.-M. Li acknowledges the support from NSFC under Grant No. 12474050 
and the MOST of China under Grant No. 2022YFA1204700. 
K. Chang acknowledges the support from NSFC under Grant No. 12488101. 
Y. Huang  acknowledges the support from NSFC under Grant No. 12304107 
and the key research and development project of Ningxia under Grants No. 2022BSB03095.

\end{document}